\documentclass[11pt,twoside]{article}
\usepackage{newpasp,graphics}
\def\edcomment#1{\iffalse\marginpar{\raggedright\sl#1\/}\else\relax\fi}
\reversemarginpar

\begin{document}
\title{Two types of radio galaxies:  a new approach}
\author{Jean Eilek}
\affil{New Mexico Tech, Socorro, NM, USA}

\begin{abstract}
We do not fully understand the dynamics and evolution of a radio
galaxy.  Models of classical double (Type II) sources are in a
reasonable state, but these objects are rare.  Non-type II sources
(generically called Type I) are far more common, but much less well
understood. In this paper I use the data to suggest possible new
models for Type I sources, and discuss the physical questions which
these new models raise.  
\end{abstract}

\section{Introduction}

What is a Type I radio source?  This question can be answered
in terms of the data (what does it look like?), or in terms of the
dynamics (how does it work?).  In neither case is the answer clear.  The
observational answer has changed over the years.  The initial Fanaroff-Riley
definition combined morphology (inner hot spots, close to the galaxy)
with low radio power ($P_{rad}$).
Later, Ledlow and Owen (1996) complicated the issue by showing that
 the division in $P_{rad}$ between Types I and II depends on the 
luminosity of the parent galaxy, $L_{opt}$.  Now, with the advent of
large radio samples, it has become clear that the morphology issue is
also more complicated than was initially realized.  When we study
images of radio galaxies which do 
not have strong outer hot spots (that is, which would not be called Type
II), we soon discover these sources are not a uniform set. 

The dynamical definition of a Type I is equally unclear. We can pose
several questions which should be answered by a dynamical model of
radio sources.  How does a given source evolve?  What governs its
size, morphology, and radio power?  How do these quantities 
change with age?  We have a reasonable picture for Type II sources, which
in principle allows quantitative answers to these questions.  We do
not, however, have a similarly useful picture for non-Type II sources, 
even though they constitute most of the radio source population.  This
is my concern in this paper. As a shorthand, I will refer to any
source which is not clearly a Type II, as a Type I.   In what follows
I will discuss  the morphology and dynamics of these interesting
sources, combining the material from my talk and poster.   

\section{Dynamical Models:  Type II sources}

Classical double sources have inspired quite a bit of study.  As a
result, we have a fairly good picture of their dynamics. A
collimated, supersonic jet (with initial velocity $v_j$, carrying
momentum flux $\Pi_j$ and energy flux $P_j$) is driven out  from the
galactic core.  It propagates into the ambient gas at a rate 
determined by its momentum flux, cross sectional area $A$ and the
density of the ambient gas, $\rho_x$ (eg, Scheuer 1974).  One way to
write this is  
\begin{equation}
\Pi_j = \rho_x A \left( { d D \over dt} \right)^2 
\end{equation}
noting that $\Pi_j \simeq P_j / v_j$, and $D$ is the length of the
jet. The resultant
propagation speed, $ d D / dt$,  is subsonic relative to the jet
material, so that the jet plasma passes through a terminal shock (identified
with the hot spots).  The post-shock jet plasma collects in a lobe/cocoon;
much of the energy carried out in the jet ends up stored in this lobe.
The lobe development is governed primarily by energy conservation (e.g.,
Eilek \& Shore 1989):
\begin{equation}
{ d E \over dt} = P_j - p { d V \over dt}
\end{equation}
if $E$, $p$ and $V$ are the internal energy, pressure and volume of the
lobe.  It is important to note here that the lobe and jet must be
treated separately;  the large volume of the lobe is needed to store
the energy ($\int P_j dt$) which has been carried out by the jet over
the lifetime of the source.  This simple model is of course modified
in detail by such effects as shocks in the ambient gas, or flows and
turbulence in the lobe,  but the global picture should still be 
accurate.  This picture has been revisited by several groups lately (eg,
Daly 1994; Kaiser, Dennett-Thorpe \& Alexander, 
1997;  Blundell, Rawlings \& Willott, 1999).

\section{What about Type I sources?}

The situation is not nearly as good for Type I sources.  The idea
arose in early work that Type I sources are turbulent, entraining
plumes (eg, De Young 1981, Bicknell 1986).   This picture was
initially attractive, and did seem a good guess for the few sources
which had been well studied at that point, such as 3C31.  
Various authors developed detailed models based on this cartoon,
assuming a steady, one-dimensional flow, and attempted to quantify
flow speeds, mass entrainment rates, or spectral aging.\footnote{Much
of this modelling has addressed only the inner regions of the sources,
where the jets are bright enough to be well imaged, and thus has not
considered more global questions such as the growth or evolution of
the entire source.} 
However,  with more data it is becoming clear that these models do not
tell the entire story.  Three clues are pointing us toward the true
situation.

One clue comes simply from the appearance
of high-quality images. Examples include the tailed sources imaged by
O'Donoghue, Owen \& Eilek (1990). The sources often show complex internal 
structures --filaments -- aligned with the overall
flow.  They can show sudden, dramatic changes in the flow, changing in
a few kpc from a well-collimated jet to a broader, poorly collimated
tail.  These common features do not seem related to terrestrial
examples of subsonic, turbulent plumes (although it may be that
turbulent flows of a magnetized plasma in the complex atmosphere
surrounding a radio galaxy may lead to such structures).  Sources
which appear easily to fit the turbulent-plume picture, such as 3C31,
turn out to be rare (cf. Parma et al 1999, or Owen \& Ledlow 1997).  
 
Another
clue comes from spectral index imaging.  Examples include 3C449
(Katz-Stone \& Rudnick 1996), 3C465 (Owen \& Eilek unpublished),  and 
two tailed sources imaged by Katz-Stone et al (1999). 
 These data clearly reveal
two-dimensional structure, in which flat-spectrum ``jets'' can be
found within steep-spectrum ``tails'' or ``sheaths''.  As flatter
spectra are associated with some combination of younger plasma and
higher magnetic fields, these images suggest that much more is taking
place than simple, one-dimensional turbulent flow.  

A third clue comes from
trying to apply equations (1) and (2) to a turbulent-plume model
(Eilek, unpublished).  Successful use of these equations (with
reasonable estimates of jet power and speed) requires the
energy-storage volume to be much larger than that of the propagating
jet.  This is, however, inconsistent with existing models of turbulent
plumes. While such simple modelling is clearly less robust than good 
data, this result again suggests that a Type I ``tail'' is better
modelled as two-dimensional , with a ``core'' and a ``sheath''. 

\section{Type I sources:  What do the Data Tell Us?}

In my opinion, we are far from a clear understanding of how Type I
sources work.  To gain insight,  my colleagues and I have been working
with a new data set, the Ledlow-Owen sample of radio sources in Abell
clusters.  We initially expected that the evolution of a source would
be determined by such factors as its jet power, parent galaxy size, or
the density and pressure of the surrounding medium.  To test these
ideas, we gathered   radio data (power and flux size), optical data
(parent galaxy magnitude) and X-ray data (from the ROSAT All-Sky
Survey for a subset of the sample; to obtain projected distance from
X-ray peak and estimated cluster gas density local to the radio
source).  The data and results are summarized in Eilek et al (2000),
and will be presented in more detail in Eilek et al (2001, in
preparation).

My particular role in this project was to measure the flux sizes of
each of the sources, and (with a grad student, T. Markovic) to
determine the source's relation to the X-ray peak (which is a good
measure of the dynamical center of the cluster).  Working individually
with images over 200 sources taught me a great deal about the nature
of the sources, and led me to an unexpected conclusion.  That
is:  although almost all of these sources would be Type I based on
their $(P_{rad}, L_{opt})$ values, they are clearly not a uniform set.
Of the 197 sources which were well-enough resolved to estimate their
structure, 188 could easily be put into one of two classes.  Our
classification criterion is, {\it how far does the jet propagate
undisturbed?}  An example of each class is shown in Figure 1.  The
two classes are as follows. 

\begin{figure}[htb]
\begin{minipage}{0.60\textwidth}
 \resizebox{\textwidth}{!}{\includegraphics{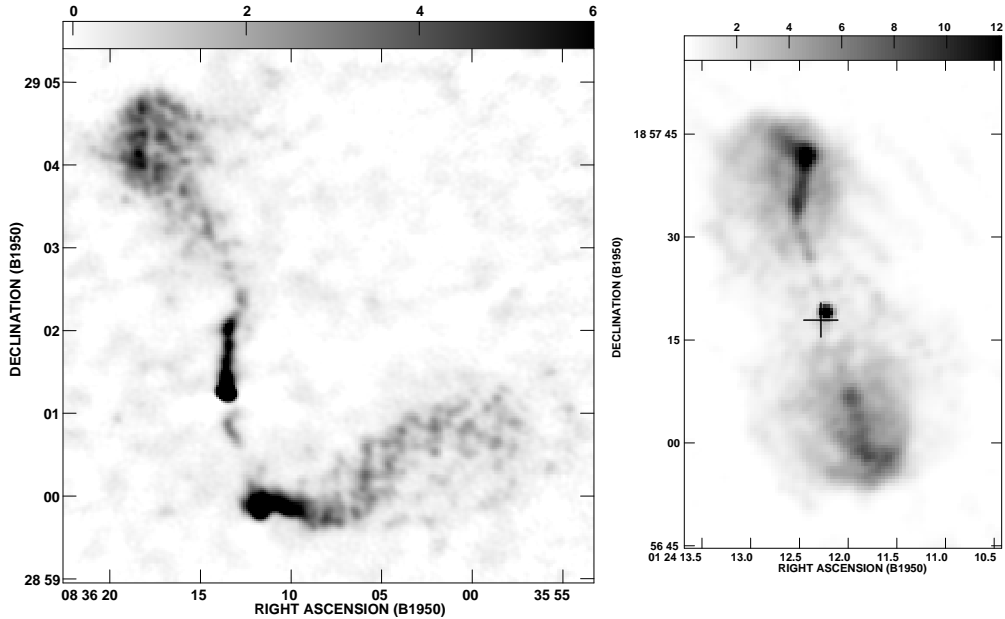}}%}
\end{minipage}
\begin{minipage}{0.40\textwidth}
\resizebox{\textwidth}{!}{\includegraphics{Eilek_fig1b.ps}}%}
\end{minipage}
\caption[]{Left,  0836+290, an example of a tailed source. The
transition  from a narrow inner jet, through an interior hot spot, to
a  tailed flow, is apparent. Right, 0124+189, an
example of a straight source.  The inner jet can be seen within each
lobe;  the jet in the north lobe stays brighter and forms a
weak hot spot.} 
\end{figure}

\begin{itemize}

\item
{\bf Straight sources} comprise about 1/3 of the set.  In these
sources, the jet retains its identity all the way to the outer end of
the source, where it may or may not end in an outer bright spot.  The
end of the flow (what might be called a working surface) is apparent
in most of the images.  The jet is generally embedded in a lobe or
cocoon (although this may be hard to detect in some of the fainter
sources, leading to what others have called a ``naked jet'').  The
sources as a class are not strongly bent;  the degree of jet
collimation varies across the class.  Classical Type II sources fall
into this category;  there are 13 of them in the Ledlow-Owen sample.

\item

{\bf Tailed sources} comprise about 2/3 of the set.  In these sources,
the jet begins narrow and well-collimated.  It soon undergoes a
dramatic transition at an inner hot spot, where the flow becomes
broader and brighter.  High-resolution images show complex structure
within these hot spots.  The flow does not fully disrupt at the hot
spot, but continues on into a radio tail.  In some
sources the end of the tail can be seen (at low frequencies), while in others
the surface brightness fades too fast to allow detection of the
end. Sources which have been called Wide-Angle or Narrow-Angle Tails
fall into this category.  However, a large number of these sources do
not bend at the hot spots;   bending is not a defining property of
the group.

\end{itemize}

This morphological division correlates nicely with the work of Parma
et al (1999).  They found that Type I sources (from the B2 sample) divide
into two spectral types:  those in which the spectrum steepens going
away from the core, and those in which it flattens.  Based on
available spectral information (usually one-dimensional; for instance
from O'Donoghue etal 1990, or comparing the sources used by Parma et
al), we find  
that that Parma et al have identified the same groups.  Tailed sources
tend to have spectral steepening outwards, away from the (interior)
jet/tail transition, while straight sources tend to have spectra
steepening inwards, away from the (exterior) end of the flow.

What determines whether a given radio galaxy will be Straight or Tailed?
We were disappointed to find no correlation between the two classes and
anything we have been able to measure (Eilek et al 2000).  
The two classes do not differ in radio power;  thus they do not differ in
underlying jet power, unless the conversion efficiency to radio power is
quite different in the two classes.  They do not differ significantly 
in  local gas density, offset from the cluster X-ray center,
 or magnitude of the parent galaxy.  They do not differ in radio
size, so the difference between them is unlikely to be due to one evolving
into the other.   There is no evidence of unusual conditions in the
ambient cluster gas (such as a discontinuity) at the hot spots in
the Tailed sources.  	It follows that there must be an internal
discriminant, such as jet density, velocity or magnetic field, which
determines the nature of the source.  I expand on this in \S 5.3.

To close this section, I note that the radio sources in the Ledlow-Owen
sample are representative of the general (nearby)
radio source population.  The fact that
they lie in Abell clusters does not make them atypical.  For instance,
Miller et al (1999) worked with a set of nearby radio galaxies (both
Types I and II) not in Abell clusters.  They combined optical and
radio data to show that the sources lie in a low-richness extension of
the Abell clusters (with some tendency for Type I's to be more
frequently associated with extended X-ray emission).  Also, Ledlow,
Owen \& Eilek (2000) compared optical and radio properties of the
Ledlow-Owen sources (within Abell clusters) to those of a set of B2
and Wall-Peacock sources (not within Abell clusters).  They found no
significant differences between the two sets in optical or radio 
properties.  Thus, there seems to be no ground for considering radio
sources in Abell clusters to be unusual.  Conclusions drawn from them
can safely be extrapolated to the general population. 

\section{Type I Sources:  What of the Physics?}

In the previous section I argued that Type I radio galaxies can be divided
into two classes, Straight and Tailed, based on their morphology.  This
new view leads us to new physical questions which must be posed (and
eventually answered).  Three questions occur to me, for which I offer
answers of varying robustness.

\subsection{What picture might work for the Straight sources?}

I suspect that all Straight sources can be modelled in terms of their
global evolution  just as classical doubles are. In fact, classical
doubles can be viewed simply as a subset of this group.  
That is, all of these sources obey momentum and energy conservation through
their length and lobe volume, as described in \S 2.  If this is the
case, then their modelling should be straightforward.  Predictions such
as evolution in the $(P_{rad}, D$) plane can be made and tested much
as has been done for Type II's.  Details of the predictions,
particularly the luminosity, will need to be modified for those
Straight sources with less well-collimated jets and fainter, or
absent, exterior hot spots.   The embedded jets in these sources
are very likely to be flatter spectrum than the lobes which surround
them (although I am not aware of two-dimensional spectral imaging on
any non-Type II straight sources).  

\subsection{What picture might work for the Tailed sources?}

I also suspect that Tailed sources involve more complicated flows. 
The transition from jet to tail, which defines these sources, 
is very suggestive of an instability which suddenly grows and
saturates. Spectral data suggests that these
tails also have a two-dimensional, jet-plus-sheath, structure.
Models similar to those proposed for Straight sources may be the place
to begin. However, the development of these sources  may be more
sensitive to external conditions than is the case for the straight
ones.  Possible complications are flows 
in the ambient medium, turbulence and mixing, and bouyancy.  Note that
Tailed sources which bend, usually do so at the hot spot.  Flows in
the ambient medium may be responsibile for the bending and may also
extend the length of the tail.  Note, also, the 
broader, ``cap-like'' ends which can be seen in some tailed sources are
reminiscent of neutral buoyancy  flows (e.g., Churazov etal
2000, in a different context).  The wide range of behavior found in
this group, from unbent 
sources to Narrow-Angle Tails, may make this set difficult to treat
uniformly or statistically.  Case studies of several individual
sources may be more profitable here.

\subsection{Why are there two types of Type I's?}

My colleagues and I
 were unable to find any measured quantity (radio, optical or X-ray)
which discriminates between Tailed and Straight sources.  
If the answer is not in something we can measure, it must be in something
we cannot measure.  That is, there must be an internal variable which differs
between the two classes.  If the Tailed sources are the result of an
instability which suddenly grows and then saturates, then the difference 
between the two classes must be the onset, or not, of this instability.  
It would follow that all sources start life looking ``straight'', that is,
with a jet propagating into the local ambient medium, depositing a cocoon
as it goes.  If the jet is unstable, nonlinear growth sets in before the
jet has gone further than several tens of kpc, and the post-instability
flow gives the source a Tailed morphology.  If the jet is stable, it
continues as it has been, growing in length and leaving excess jet material
behind in a lobe.  

It remains is to determine what sets off the instability. If we
consider a Kelvin-Helmholtz type of instability, we know from analytic
and numerical methods (e.g., Rosen et al
1999 and references therein) that it is sensitive to the density, speed and
magnetic field of the flow (see Eilek et al 2000 for more discussion). A
change in any of these quantities (for instance as the jet expands or
the cocoon evolves) has the potential to set off the instability.  It
also remains to determine whether such an instability can grow and saturate
quickly enough to reproduce the hot spots we observe.  This is probably
best tested by numerical simulations;  until such time this
saturated-instability model is  only a cartoon.

\section{Caveats on Spectral Aging}

One important application of a dynamical model of radio source evolution
is to the interpretation of spectral data.  Spectral steepening is
commonly interpreted  
as arising from aging of the electron population due to synchrotron losses.
It has often been used as a measure of the age of a radio
source. Several authors have carried out such analyses for Type II
sources (e.g., Leahy, Muxlow \& Stephens 1989, Carilli et al 1991) and
a few have done so for Type I's (e.g., Parma et al 1999).  
If we have a good dynamical model of a radio source, we can (in principle,
anyway) determine both its physical age and the internal distribution of 
its magnetic field with some confidence.  We can then use this
information to track the 
emission history of the relativistic electrons, and thus test the hypothesis
that the spectral steepening is due to electron aging.  

In practice, of course, such analyses are hampered by necessary but
probably unsupported assumptions.  Examples of this include
assumptions about the magnetic field behavior,  about the 
shape of the spectrum (which must be inferred from a small number of
frequencies), or about the relation between hot spot equipartion pressure and
jet  thrust (see also comments by  Blundell \& Rawlings 2000).
A further complication is now apparent:  the sources are
multi-dimensional.  That is, neither surface brightness {\it nor spectrum} 
can be described as one-dimensional functions of distance away from the
hot spot (whether inner or outer).  This is true to some extent of Type II
sources (e.g., Leahy et al 1989), and is particularly true of Type I
sources (such as those mentioned in \S 3).  The two-dimensional complexity
of the spectral images can vitiate conclusions based on a one-dimensional
``slice'' analysis.  As an example (illustrated by data in Katz-Stone
et al 1999), consider a tailed source in which
flat and steep spectrum components coexist.  If the relative mix of
these two components changes, such as the steep spectrum component
suddenly becoming brighter,  a ``slice'' analysis will erroneously
conclude that the flow has suddenly decelerated -- when this is not at
all the true situation.

\section{Conclusions}
All  I can really conclude is that Type I sources remain a
fascinating challenge to our understanding of radio galaxy evolution.
I anticipate that combining high-quality imaging with new theoretical
analysis will improve our understanding,  and no doubt expose the
naivety of my speculations here.

\begin{acknowledgements}
My understanding of radio galaxy physics has benefitted immensely from
ongoing conversations with colleagues such as Dave De Young, Phil
Hardee, Robert Laing, Aileen O'Donoghue, Frazer Owen, and Larry
Rudnick.   This work was partially supported by NSF grant AST-9720263.  
\end{acknowledgements}


\begin{references}

\reference
Bicknell, G. V., 1986, 'A model for the suface brightness of a
turbulent, low Mach number jet.  II - the global energy budget and
radiative losses', \apj, 300, 591-604

\reference
Blundell, K. M., Rawlings, S. \& Willott, C. J., 1999, 'The nature and
evolution of classical double radio sources from complete samples',
\aj, 117, 677-706

\reference
Blundell, K. M. \& Rawlings, S. 2000, 'The spectra and energies of
classical double radio lobes', \aj, 119, 1111-1122

\reference
Carilli, C. R., Perley, R. A., Dreher, J. W. \& Leahy, J. P. 1991,
'Multifrequency radio observations of Cygnus A:  spectral aging in 
powerful radio sources', \apj, 383, 554-573

\reference
Churazov, E., Forman, W. Jones, C. \& B\"ohringer, H. 2000, 'Asymmetric,
arc-minute scale structures around NGC 1275', \aap, 356, 788-794.
 
\reference
Daly, R., 1994, 'Cosmology with powerful extended radio sources',
\apj, 426, 38-50

\reference
De Young, D. S., 1981, 'Emission line regions and stellar associations
in extended extragalactic radio sources', \nat, 293, 43-44

\reference
Eilek, J., Hardee, P., Markovic, T., Ledlow, M. \& Owen, F. 2000, 'On
dynamical models for radio galaxies', to appear in New Astronomy
Reviews, Life Cycles 
of Radio Galaxies, ed. J. Biretta et al

\reference
Eilek, J. A. \& Shore, S. N. 1989, 'The energetics and evolution of jet-fed
radio sources', \apj, 342, 187-207

\reference
Katz-Stone, D. M. \& Rudnick, L. 1996, 'An analysis of the synchrotron
spectrum in the Fanaroff-Riley Type I galaxy 3C449', \apj, 488, 146-154

\reference
Katz-Stone, D. M., Rudnick, L., Butenhoff, C. \& O'Donoghue, A. A.
1999, 'Coaxial jets and sheaths in wide-angle tailed radio galaxies',
\apj, 516, 716-728

\reference
Kaiser, C. R., Dennett-Thorpe, J. \& Alexander, P. 1997, 'Evolutionary
tracks of FRII sources through the P-D diagram', \mnras, 292, 723-732

\reference
Leahy, J. P., Muxlow, T. W. B. \& Stephens, P. W., 1989, '151-MHz and
1.5-GHz observations of bridges in powerful extragalactic radio
sources', \mnras, 239, 401

\reference
Ledlow, M. J. \& Owen, F. N. 1996, '20 cm VLA survey of Abell clusters
of galaxies.  VI.  Radio/optical luminosity functions', \aj, 112, 9-22

\reference
Ledlow, M., J., Owen, F. N. \& Eilek, J. A. 2000, 'Rich cluster and
non-cluster radio galaxies \& the (P,D) diagram for a large number of
FRI and FRII sources', to appear in New Astronomy Reviews, Life Cycles
of Radio Galaxies, ed. J. Biretta etal

\reference
Miller, N, A., Owen, F. N., Burns, J. O., Ledlow, M. J., \& Voges, W.
1999, 'An X-ray and optical investigation of the environments rround
nearby radio galaxies', \aj, 118, 1988-2001

\reference
O'Donoghue, A. A., Owen, F. N. \& Eilek, J. A. 1990, 'VLA observations of
Wide-Angle Tailed radio sources', \apjs, 72, 75-131

\reference
Owen, F. N. \& Ledlow, M. J. 1997, 'A 20 cemtimeter VLA survey of
Abell clusters of galaxies.  VII.  Detailed radio images', \apjs, 108,
41-98

\reference
Parma, P., Murgia, M., Morganti, R., Capetti, A., de Ruiter, H. R. \&
Fanti, R. 1999, 'Radiative ages in a representative sample of low
luminosity radio galaxies', \aap, 344, 7-16

%\reference
%Parma, P., Murgia, M., de Ruiter, H. R. \& Fanti, R. 2000, 'The Lives
%of FR I Radio Galaxies', to appear in New Astronomy Reviews, Life Cycles
%of Radio Galaxies, ed. J. Biretta etal

\reference
Rosen, A., Hughes, P. A., Duncan, G. C. \& Hardee, P. E., 1998, 'A
comparison of the morphology and stability of relativistic and 
nonrelativistic jets', \apj, 516, 729-743

\reference
Scheuer, P. G., 1974, 'Models of extragalactic radio sources with a
continuous energy supply from a central object', \mnras, 166, 513-528
\end{references}
\end{document}